\documentclass[sigconf,authorversion,nonacm]{acmart}

\AtBeginDocument{%
	}

\usepackage{subfigure}
\usepackage{multirow}
\usepackage{tabularx}
\usepackage{makecell}
\pdfoutput=1

\begin{document}
	
	\title{Machine Learning Techniques for Python Source Code Vulnerability Detection}
	
	\author{Talaya Farasat}
	
	\affiliation{%
		\institution{University of Passau}
		\streetaddress{Passau, Germany}
		\city{Passau}
		\country{Germany}}
	
	\author{Joachim Posegga}
	
	\affiliation{%
		\institution{University of Passau}
		\streetaddress{Passau, Germany}
		\city{Passau}
		\country{Germany}}
	
	\renewcommand{\shortauthors}{Farasat and Posegga}
	
	\begin{abstract}
		Software vulnerabilities are a fundamental reason for the prevalence of cyber attacks and their identification is a crucial yet challenging problem in cyber security. In this paper, we apply and compare different machine learning algorithms for source code vulnerability detection specifically for Python programming language. Our experimental evaluation demonstrates that our Bidirectional Long Short-Term Memory (BiLSTM) model achieves a remarkable performance (average Accuracy = 98.6\%, average F-Score = 94.7\%, average Precision = 96.2\%, average Recall = 93.3\%, average ROC = 99.3\%), thereby, establishing a new benchmark for vulnerability detection in Python source code.
	\end{abstract}

	
	

	\maketitle
	
	\section{Introduction}
	Code flaws or vulnerabilities are prevalent in software systems and can potentially lead to system compromise, information leaks, or denial of service. 
	Recognizing the constraints of traditional methods (static \& dynamic code analyses), and with the growing accessibility of open-source software repositories, it has been recommended to adopt a data-driven approach for software vulnerability detection. Therefore,  various machine learning techniques have been applied to learn vulnerable features of source code, and to automate the process of software vulnerability identification\cite{b1, b4, b5, b6, b14, b16, b18, b22}. Many researchers focus on source code vulnerability detection across various programming languages such as Java, C, and C++. Some notable studies include \cite{b1, b5, b9, b10, b21, b22}. In 2024, Python continues to maintain its prominent position as one of the top programming languages \cite{b41}, and also a majorly used language on GitHub \cite{b39}. Despite its popularity, Python has been relatively overlooked by researchers. Only a few studies \cite{b6, b14, b18, b24} focus on vulnerability detection specifically in Python programming language.
	
	Given the abundance of machine learning algorithms and their corresponding hyper-parameters available, there is potential for improved results in this domain. To bridge this gap, we apply and compare five different machine learning models. Notably, our BiLSTM model demonstrates superior performance as compared to all other applied models and also with the approaches presented in \cite{b6, b14, b18}.
	We also open-source all our code and models used in this study for broader dissemination. Our code and
	models can be accessed at https://github.com/Tf-arch/Python-Source-Code-Vulnerability-Detection/tree/main
	\begin{table*}	
		
		\renewcommand{\arraystretch}{0.9}
		\begin{tabular}{|p{3.1cm}|c|c|c|c|c|c|c|c|c|c|c|}
			\hline
			\multirow{2}{3.9cm}{Vulnerabilities} & \multicolumn{2}{c|}{\textbf{GNB}} & \multicolumn{2}{c|}{\textbf{Decision Tree}} & \multicolumn{2}{c|}{\textbf{LR}} & \multicolumn{2}{c|}{\textbf{MLP}} & \multicolumn{2}{c|}{\textbf{BiLSTM}}\\
			\cline{2-11}
			& Accuracy & F-Score & Accuracy & F-Score   & Accuracy & F-Score & Accuracy & F-Score & Accuracy & F-Score\\
			\hline
			SQL injection & 81.0\% & 0.63\% & 80.5\% & 0.97\% & 83.7\% & 51.2\% & 87.6\% & 65.3\% & \textbf{98.2\%} & \textbf{95.3\%}\\ \hline
			
			XSS & 8.8\% & 16.0\% & 91.5\% & 15.8\% & 94.8\% & 68.9\% & 95.4\% & 72.6\% & \textbf{98.8\%} & \textbf{93.0\%}\\ \hline
			
			Command injection & 71.8\% & 24.5\% & 86.1\% & 3.6\% & 96.0\% & 83.7\% & 93.1\% & 70.4\% & \textbf{99.1\%} & \textbf{96.7\%}\\ \hline

			XSRF & 14.2\% & 23.6\% & 86.1\% & 1.8\% & 89.6\% & 56.5\% & 93.9\% & 76.8\% & \textbf{98.3\%} & \textbf{93.6\%}\\ \hline
			
			Remote code execution & 9.5\% & 15.9\% & 90.7\% & 0.74\% & 98.2\% & 89.5\% & 98.9\% & 93.8\% &\textbf{99.4\%} & \textbf{96.5\%}\\ \hline

			Path disclosure & 11.8\% & 20.8\% & 85.8\% & 5.1\% & 95.8\% & 81.3\% & 97.6\% & 89.8\% & \textbf{99.3\%} & \textbf{97.3\%}\\ \hline

			Open redirect  & 14.4\% & 23.7\% & 86.6\% & 1.1\% & 86.7\% & 48.9\% & 89.0\% & 58.0\% & \textbf{97.5\%} & \textbf{90.7\%}\\ \hline

		\end{tabular}
		\caption{Performance Evaluation of Machine Learning Algorithms}
		
	\end{table*}	
	\begin{figure*}
		\centering
		\subfigure[SQL injection ROC]{
			\includegraphics[width=39mm, height=3.5cm]{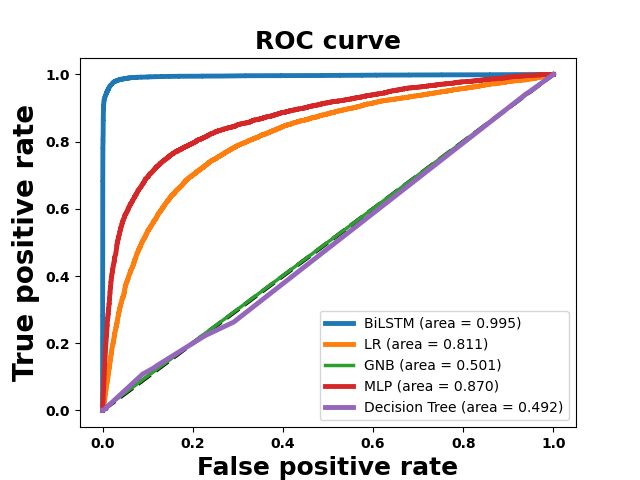}
		}
		\subfigure[XSS ROC]{
			\includegraphics[width=39mm, height=3.5cm]{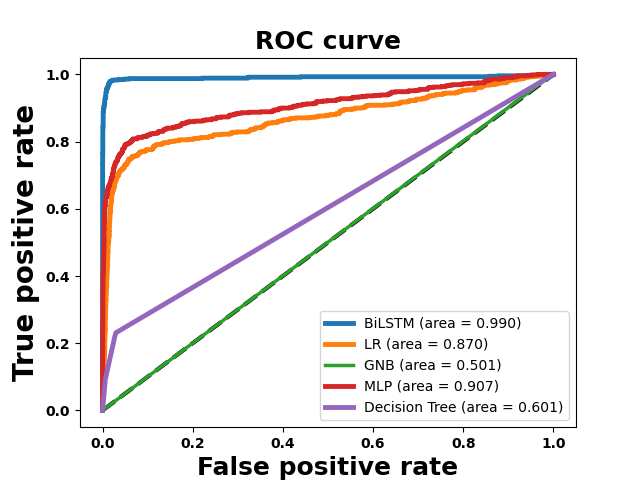}
		}
		\subfigure[Command injection ROC]{
			\includegraphics[width=38mm, height=3.5cm]{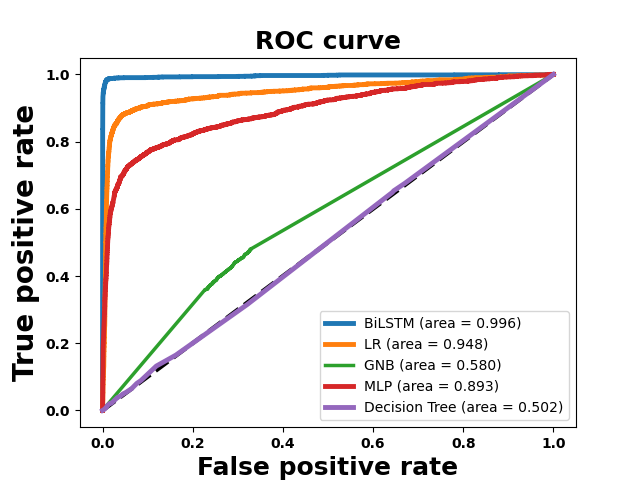}
		}
		\subfigure[XSRF ROC]{
			\includegraphics[width=38mm, height=3.5cm]{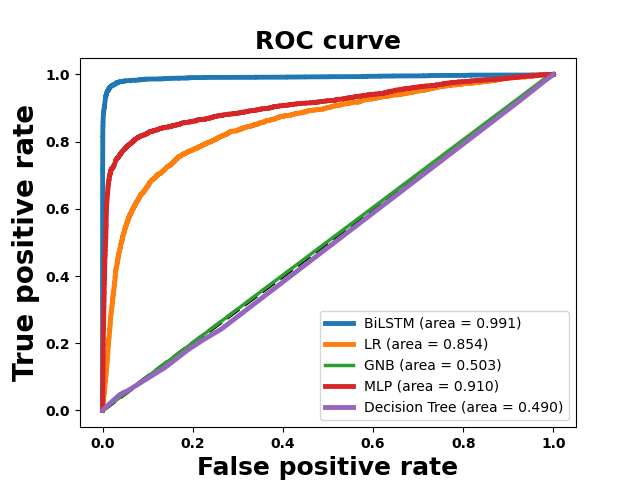}
		}

		\caption{ROC Curves (see remaining ROC results here: https://github.com/Tf-arch/Python-Source-Code-Vulnerability-Detection/tree/main)}
		
	\end{figure*}
	
	{\tabcolsep=5pt\def\arraystretch{1.4}
		\begin{table*} 
			\renewcommand{\arraystretch}{0.9}

			\begin{tabular}{|p{2.2cm}|p{2cm}|p{2.2cm}|p{2.2cm}|p{2.3cm}|c|c|c|c|c|c|}
				\hline
				\multirow{1}{3cm}{\textbf{Vulnerabilities}} & \multirow{1}{1.2cm}{\textbf{Metrics}} & \multicolumn{1}{c|}{\textbf{Bagheri and Hegedűs\cite{b18}}}& \multicolumn{1}{c|}{\textbf{Wartschinski et al. \cite{b6}}} & \multicolumn{1}{c|}{\textbf{Wang et al. \cite{b14,b45}}} & \multicolumn{1}{c|}{\textbf{Our Model}}\\
				\cline{3-5}
				
				\hline{}
				& Precision & 82.2\% & 82.2\%  & 84.4\%  & \textbf{96.8\%}  \\ \cline{2-6}
				\textbf{SQL injection}& Recall & 78.0\% &78.0\%  & 73.9\%  & \textbf{93.8}\%   \\ \cline{2-6}
				& F-Score & 80.1\% & 80.1\%  & 78.8\%  & \textbf{95.3\%} \\ \cline{2-6}
				& Accuracy & 92.5\% & 92.5\%  & - & \textbf{98.2\%} \\ \cline{1-6}
				
				& Precision & 91.9\%& 91.9\%  & 97.0\%  & \textbf{94.8\%}  \\ \cline{2-6}
				\textbf{XSS}& Recall & 80.8\% & 80.8\%  & 79.7\%  & \textbf{91.3}\%   \\ \cline{2-6}
				& F-Score & 86.0\%  & 86.0\%  & 87.5\%  & \textbf{93.0\%} \\ \cline{2-6}
				& Accuracy & 93.8\% & 97.8\%  & - & \textbf{98.8\%} \\ \cline{1-6}

				& Precision & 94.0\% & 94.0\%  & 92.5\%  & \textbf{97.8\%}  \\ \cline{2-6}
				\textbf{Command}& Recall & 87.2\% & 87.2\%  & 85.2\%  & \textbf{95.7}\%   \\ \cline{2-6}
				\textbf{injection}& F-Score  & 90.5\% & 90.5\%  & 88.7\%  & \textbf{96.7\%} \\ \cline{2-6}
				& Accuracy & 95.8\% & 97.8\%  & - & \textbf{99.1\%} \\ \cline{1-6}

				& Precision & 92.9\%& 92.9\%  & 88.0\%  & \textbf{96.7\%}  \\ \cline{2-6}
				\textbf{XSRF}& Recall & 85.4\% & 85.4\%  & 80.5\%  & \textbf{90.7\%}   \\ \cline{2-6}
				& F-Score & 89.0\% & 89.0\%  & 84.0\%  & \textbf{93.6\%} \\ \cline{2-6}
				& Accuracy & 92.2\% & 97.2\%  & - & \textbf{98.3\%} \\ \cline{1-6}

				& Precision & 96.0\%  & 96.0\%  & 93.5\%  & \textbf{97.2\%}  \\ \cline{2-6}
				\textbf{Remote code}& Recall & 82.6\% & 82.2\%  & 77.2\%  & \textbf{95.9}\%   \\ \cline{2-6}
				\textbf{execution}& F-Score & 88.8\% & 88.8\%  & 84.6\%  & \textbf{96.5\%} \\ \cline{2-6}
				& Accuracy & 91.1\% & 98.1\%  & - & \textbf{99.4\%} \\ \cline{1-6}

				& Precision & 92.0\% & 92.0\%  & 90.6\%  & \textbf{97.7\%}  \\ \cline{2-6}
				\textbf{Path disclosure}& Recall  & 84.4\% & 84.4\%  & 82.7\%  & \textbf{96.9}\%   \\ \cline{2-6}
				& F-Score & 88.1\% & 88.1\%  & 86.5\%  & \textbf{97.3\%} \\ \cline{2-6}
				& Accuracy & 91.3\% & 97.3\%  & - & \textbf{99.3\%} \\ \cline{1-6}

				& Precision& - & 91.0\%  & 80.8\%  & \textbf{92.5\%}  \\ \cline{2-6}
				\textbf{Open redirect}& Recall & - & 83.9\%  & 84.3\%  & \textbf{89.0}\%   \\ \cline{2-6}
				& F-Score & - & 87.3\%  & 82.5\%  & \textbf{90.7\%} \\ \cline{2-6}
				& Accuracy  & - & 96.8\%  & - & \textbf{97.5\%} \\ \cline{1-6}

			\end{tabular}
			\caption{Comparison of our work with related work}
			
		\end{table*}
		\section{Experimental Design}
		We're examining the same software vulnerabilities highlighted in \cite{b6}, \cite{b14}, and \cite{b18}, i.e., SQL injection, cross-site scripting (XSS), command injection, cross-site request forgery (XSRF), path disclosure, remote code execution, and open redirect. 
		
		\textbf{Dataset:} We use the dataset prepared by Wartschinski et al. \cite{b6}, available at \cite{b42}, which is compiled by targeting publicly accessible GitHub repositories. GitHub stands out as the largest repository hosting platform for source code globally, making it an ideal resource for this work.
		Wartschinski et al. \cite{b6} gather a distinct dataset for each vulnerability type. The data is collected in the form of commits that contain security-related fixes. Sections of code that are updated or removed in these commits are categorized as vulnerable, along with the surrounding code to provide context. Conversely, the remaining code and the post-fix version are labeled (probably) as not vulnerable. We use 70\% data in training, 15\% in testing, and 15\% in the validation set.
		
		\textbf{Word2vec Embeddings:}
		For the training of machine learning algorithms, it is necessary to represent code tokens as vectors that retain the semantic and syntactic information. We train our word2vec model with the help of instructions outlined here \cite{b42}. We use the same hyper-parameters for the word2vec model highlighted in \cite{b6}, i.e., training iterations: between one and more than a hundred, vector dimensionality: between 5 and 300, minimum count: between 10 and 500. We test different models with our machine-learning algorithms. We get the best results with training iterations: 200, minimum count: 10, and vector dimensionality: 300. Therefore, we use this word2vec model (10,200,300) with the majority of machine learning algorithms.

		\textbf{Machine Learning Algorithms:}
		We utilize the Python sci-kit-learn library to construct Gaussian Naive Bayes (GNB), Decision Tree, Logistic Regression (LR), and Multi-Layer Perceptron (MLP). For GNB, we employ default parameters.
		In the case of Decision Tree, default parameters are used, with the exception of setting the max\_depth parameter. Specifically, we set max\_depth as 2 for XSS and open\_redirect vulnerabilities, while for other vulnerabilities, we use a max\_depth value of 5. For Logistic Regression, default parameters are utilized, except for the solver parameter, which is set to liblinear. Lastly, in MLP, we opt for default parameters, except for the max\_iter parameter, which is set to 300.

		\textbf{Bidirectional Long Short-Term Memory} (BiLSTM) in contrast with LSTM, processes sequential data in both forward and backward directions with two separate hidden layers. It enables additional training by traversing the input data twice. We use the Python Keras framework (backend Tensorflow) to create our BiLSTM model. 
		
		\textbf{Selection of optimal hyper-parameters for BiLSTM:} As BiLSTM networks are highly configurable through several hyper-parameters. Choosing the correct set of hyper-parameters is crucial because it directly impacts the models' performance and helps in achieving benchmark results. We can alter/tune several hyper-parameter values like the number of BiLSTM layers, optimizers, learning rate, loss functions, number of epochs, and batch sizes. We experiment with changes in the hyper-parameter values manually. As per our experimental results, we observe that the following combination of hyper-parameters achieves remarkable performance: one input layer, three hidden layers with 50 BiLSTM cells or neurons, (BiLSTM creates two copies of the hidden layer, and the output values from these BiLSTMs are concatenated), four dropout layers with 0.2 dropout rate, and
		an output layer with a single node. We use Adam optimizer and mean\_squared\_error as the loss function. The model is trained for 50 epochs with the batch size = 128.
		
		\textbf{Performance Evaluation:} We evaluate and compare the performance of machine learning algorithms based on accuracy and F-score, see Table 1. We also evaluate the Receiver Operating Characteristics (ROC) curves on the validation dataset, see Figure 1.  Experimental results show that our BiLSTM model with the optimized hyper-parameter values, can effectively detect the Python source code vulnerabilities with the highest Accuracy, F-Score, and ROC curve values (average Accuracy= 98.6\%, average F-Score= 94.7\%, average ROC= 99.3\%), see Table 1 and Figure 1. 
		\section{Related Work}
		Our work is closely related to Wartschinski et al. \cite{b6}, Bagheri and Hegedűs \cite{b18}, and Wang et al. \cite{b14}. Wartschinski et al. \cite{b6} applied LSTM (with word2vec) for Python source code vulnerability detection. Bagheri and Hegedűs \cite{b18} results show that  BERT embeddings with the LSTM model achieve the best overall accuracy in predicting Python code vulnerabilities.  Similarly, Wang et al. \cite{b14} presented the empirical study of different deep learning models for Python source code vulnerability detection. They show that LSTM  and gated recurrent unit (GRU) can indeed be the best choice. They also highlight that BiLSTM and GRU with attention (using word2vec) are two optimal models for Python source code vulnerability detection.
		
		Following their research, we apply BiLSTM with word2vec for Python source code vulnerability detection.
		Different from the studies \cite{b6, b14, b18}, we also apply and compare GNB, LR, Decision Tree, and MLP models as well, and compare their performances with each other, see Table 1 and Figure 1. Moreover, by leveraging the optimal settings of our BiLSTM model, we achieve a remarkable performance (average Accuracy= 98.6\%, average F-Score= 94.7\%, average Precision= 96.2\%, average Recall= 93.3\%, average ROC= 99.3\%), thereby, establishing a new benchmark for vulnerability detection in Python source code, see Table 2 for detailed comparison (Bagheri and Hegedűs \cite{b18} (LSTM with BERT), Wang et al. \cite{b14, b45} (LSTM with word2vec), our model (BiLSTM with word2vec)). Our results can be observed here as well: https://github.com/Tf-arch/Python-Source-Code-Vulnerability-Detection/tree/main. 
		
		Ehrenberg et al. \cite{b24} apply named entity recognition techniques for recognizing the Python source code vulnerabilities.
		We do not compare our results in detail with them, as their methodology primarily focuses on named entity recognition, differing substantially from our approach. Nevertheless, it's notable that their overall average accuracy stands at 97.29\%, with an average F1-score of 91.86\%. In contrast, our BiLSTM model yields superior performance with an average accuracy of 98.6\% and an average F1-score of 94.7\%.
		
		\section{Conclusion}
		
		In this work, we apply and compare five different machine learning models for detecting vulnerabilities in Python. Our BiLSTM model with word2vec shows remarkable efficacy (average Accuracy = 98.6\%, average F-Score = 94.7\%, average Precision = 96.2\%, average Recall = 93.3\%, average ROC = 99.3\%), surpassing not only the other machine learning models we applied but also outperforming the techniques detailed in prior works \cite{b6, b14, b18}. We believe this work is helpful for researchers and Python programmers facing daily challenges related to identifying programming vulnerabilities.




	\end{document}